# The effect of doping on the lattice parameter and properties of cubic boron nitride

Vladimir A. Mukhanov,[a] Alexandre Courac [b] and Vladimir L. Solozhenko [a,*]

[a] LSPM–CNRS, Université Sorbonne Paris Nord, 93430 Villetaneuse, France

[b] Institut de Minéralogie, de Physique des Matériaux et de Cosmochimie (IMPMC)
Sorbonne Université, UMR CNRS 7590, 75005 Paris, France

*The effect of doping of cubic boron nitride with beryllium, silicon, sulfur and magnesium on the lattice parameters, electrical conductivity and EPR spectra has been studied. It is established that the degree of doping increases significantly in the case of crystallization of cubic boron nitride from BN solutions in supercritical ammonia at 3.9-4.2 GPa and 1100°C in comparison with the conventional synthesis from melts of the Mg–B–N system at 4.2 GPa and 1400°C. Doping with silicon and beryllium results in semiconductor properties of cubic boron nitride.*

## Introduction

Cubic boron nitride (cBN), the second hardest industrial-scale material after diamond [1-3], is remarkable not only for its thermal and chemical stability significantly exceeding the stability of diamond (especially with respect to oxygen and iron alloys), but also for its larger band gap (6.4±0.5 eV [4]). In contrast to diamond, cBN doping allows obtaining both p- and n-type semiconducting crystals [5]. However, despite numerous attempts to establish the nature of structural defects in cBN and their influence on the optical and semiconducting properties of the crystals [6–12], complete clarity on these issues has not yet been achieved.

From the standpoint of practical applications, the most interesting field is the synthesis of cubic boron nitride in the Mg–B–N system by crystallization from melts of this system [13–18], as well as from solutions of graphite-like hexagonal boron nitride (hBN) and magnesium-containing compounds in a supercritical N–H fluid [19–23] at high pressures and high temperatures. In both cases, cBN crystallizes in the region of its thermodynamic stability far from the line of phase equilibrium hBN ⇌ cBN [20, 24–26]. Crystallization of cBN from boron nitride solutions in supercritical fluids begins at much lower temperatures (1000–1100°C) and pressures (2.5–3.8 GPa) compared to classical synthesis from a melt (at least 1300°C and 4.0 GPa) [20]. The latter is due to the fact that the use of supercritical fluids as solvents of boron nitride can significantly reduce the activation barrier of nucleation and crystallization [23], which, in turn, leads to a substantial increase in both the number of cBN nuclei and their growth rate. This gives reason to expect that under these conditions the degree of doping of the forming crystals will be noticeably higher than in the case of a relatively slow "quasi-equilibrium" growth from high-temperature melts of the Mg-B-N system.

The aim of this work was to study the effect of doping with beryllium, silicon, sulfur and magnesium on the structure and properties of cBN crystals obtained in the Mg–B–N system by crystallization both from melts and from solutions in supercritical N–H fluids [27].

## Experimental

Experiments on the synthesis of doped cBN from melts of the Mg–B–N system and solutions in supercritical ammonia were carried out in a toroid-type high-pressure apparatus [28]. Diagrams of the used high-pressure cells with a useful volume of up to 2 cm$^3$ are shown in Fig. 1. The pressure in a cell was determined from the phase transitions in bismuth (2.52 and 2.69 GPa) and thallium (3.67 GPa). The temperature in the cell was estimated from the electric power in accordance with the calibration curve obtained in the experiments on melting of a number of substances (Si, NaCl, CsCl, etc.) and diamond synthesis in the Ni–Mn–C system. The details of pressure and temperature calibration have been described previously [29]. The absence of significant temperature gradients in the cell was achieved by using composite electrical leads made of a mixture of graphite (20 wt%) and heat-treated kaolin, on which up to 90% of the supplied electric power was released as heat.

The synthesis in the presence of supercritical ammonia was performed under a pressure of about 4 GPa at 1000–1100°C in copper capsules with a knife seal (Fig. 1a). A capsule filled with a reaction mixture and cooled with liquid nitrogen was loaded with liquid ammonia in nitrogen vapor. The capsule was closed with a cap and placed in a high-pressure cell pre-cooled with liquid nitrogen, which was then rapidly loaded to a pressure of about 2 GPa to seal the capsule. After that, the pressure was raised to the required value within 5 min, and the electric power necessary to reach the required temperature in the cell was supplied. The heating duration was 5–10 min, after which the sample was quenched by instantaneous switching off the heating current, and after complete cooling of the high-pressure cell down to room temperature the pressure was reduced to ambient. The degree of conversion of hBN to cBN in such an experiment reached 98–100%, while it was only 30–40% in the classical synthesis at 4 GPa and 1400°C with layer-by-layer loading of hBN and magnesium-containing compounds capable of dissolving BN (MgB$_2$, Mg$_3$N$_2$, Mg$_3$(BN$_2$)$_2$, etc.) (diagram of the corresponding cell is shown in Fig. 1b). In the first case, the doping was performed by introducing small amounts of dopants (Be$_3$N$_2$, BeO, S, Si, and Si$_3$N$_4$) into the reaction mixture of ammonia with hBN and magnesium diboride MgB$_2$ or magnesium boron nitride Mg$_3$(BN$_2$)$_2$ (up to 10 wt% with respect to hBN). In the classical synthesis, the dopants were introduced into the layers of magnesium-containing compounds (solvents). The synthesis conditions and the lattice parameter values of doped cBN are given in Table.

The synthesized product was removed from the copper capsule (or graphite heater), crushed and treated for 30 min with a molten NaOH–KOH mixture (molar ratio 1:1) at 300°C to remove residual hBN, and then water was carefully added to the melt until the alkalis were completely dissolved. The precipitate was washed with water and treated with boiling aqua regia for 5 min to completely remove magnesium-containing compounds and dopants. The remaining cBN crystals were washed with water and dried at 120°C.

The synthesized samples were studied by X-ray diffraction on a Seifert MZIII powder diffractometer (CuKα radiation) in the Bragg–Brentano geometry. The goniometer was adjusted using standard samples of high-purity silicon ($a$ = 5.431066 Å) and LaB$_6$ ($a$ = 4.15695 Å). The positions of the *111*, *200*, *220*, *311* and *400* lines of cBN were determined by fitting the profile to the Pearson function (PVII). Precise determination of the lattice parameters was performed using the U-FIT software package [30], which allows one to take into account the displacement of the sample surface from the diffraction plane, the zero shift of the goniometer, and the absorption of X-rays by the sample.

Electron spin resonance (ESR) spectra of the synthesized cBN crystals were recorded on a Bruker ER 200D SRC spectrometer at room temperature.

Sintering of doped cBN powders was carried out in copper capsules in a toroid-type high-pressure apparatus at 8 GPa and 1400°C for 1–3 min; the sinters with a diameter of 3–4 mm and a height of 1–3 mm were obtained. In some cases, cBN powders were sintered with aluminum nitride (10 wt%, particle size 0.5–5 μm), which had no effect on the electrical properties of the obtained sinters, but significantly increased their mechanical stability at temperatures up to 700°C. To extract the samples after sintering, copper capsules were dissolved in nitric acid.

To measure the electrical resistivity and its temperature dependence, the sintered samples were covered with contacts made of a multicomponent composition based on titanium and silver. Current supplying electrodes (copper, nichrome, silver or platinum wire 100–300 μm in diameter) were soldered to the contacts using an IR laser (λ = 1060 nm). Similarly, platinum contacts were soldered to individual cBN single crystals. High-temperature measurements (up to 700°C) were carried out in a quartz tube with a nichrome heater. The temperature was measured using a chromel–alumel thermocouple, the junction of which was placed at a distance of 1 mm from the sample.

## Results and Discussion

The lattice parameter of undoped cBN obtained by direct conversion of high-purity pyrolytic boron nitride at high pressures and high temperatures is $a$ = 3.6153(1) Å [31]. cBN samples synthesized by other methods can have either a larger or a smaller lattice parameter. In particular, it depends on unavoidable impurities of oxygen (air, oxidized BN surface) and carbon (impurities in the starting hBN, graphite heater of the high pressure cell). Both of these impurities tend to decrease the lattice parameter. Given that the synthesis is usually performed in the absence of additives that bind oxygen and carbon, we chose the lattice parameter of cBN synthesized by crystallization from magnesium-containing solvents as the lattice parameter that corresponds to "zero" doping because of its closeness to that mentioned above.

### cBN doped with magnesium

Black cBN crystals obtained by crystallization from melts of the Mg–B–N system at 4 GPa and 1400°C have crystal lattice parameter $a = 3.6150(1)$ Å that practically equals to that of pure cBN ($a = 3.6153(1)$ Å [31]), which indicates the absence of doping of cubic boron nitride with magnesium under these conditions.

Cubic boron nitride synthesized by crystallization from BN solutions in supercritical ammonia in the presence of magnesium diboride (or magnesium boron nitride) (Fig. 2a) has a yellow color and exhibits very high electrical resistance. For the samples synthesized by this method, the increase in lattice parameter with respect to cBN obtained from the melt of the Mg–B–N system reaches 0.0014(4) Å, which is due to the incorporation of magnesium atoms into structure of the formed cBN. The amount of magnesium incorporated into cBN lattice can be estimated using Vegard's law under assumption that cubic solid solutions of $Mg_xB_{1-x}N$ composition are formed (Fig. 3a). The lattice parameter for hypothetical sphalerite-type MgN ($a = 4.987$ Å) was calculated assuming that the length of Mg–N bond is 2.160 Å, which corresponds to the average length of Mg–N bond in $Mg_3N_2$ [32]. Within this approach, the maximum magnesium content in cBN lattice can be estimated as 0.05 at%. It should be noted that incorporation of magnesium into cBN during crystallization from BN solutions in supercritical fluids is possible only in the absence of other (except Mg) dopants.

### Attempts of cBN doping of with sulfur

Attempts to obtain sulfur-doped cubic boron nitride were made only under conditions of crystallization from BN solutions in supercritical ammonia (see Table). Apparently, incorporation of sulfur atoms into lattice of the formed cBN does not occur due to the fact that sulfur reacts with magnesium compounds to form MgS sulfide, which is insoluble in the supercritical fluid. The obtained cBN crystals (Fig. 2b) had straw-yellow color (and not reddish, as reported in [7]) and exhibited high electrical resistance.

### cBN doped with silicon

The lattice parameter of cBN crystals obtained by crystallization from the melt of the BN-$Mg_3N_2$-$Si_3N_4$ system ($a = 3.6153$ Å) practically does not differ from the lattice parameter of undoped cBN, which indicates an extremely low solubility of silicon in cubic boron nitride under conditions of the classical synthesis. Nevertheless, the pronounced brown color and noticeable electrical conductivity (about $10^{-3}$ $\Omega^{-1}$) of the obtained crystals unambiguously indicate the incorporation of silicon into cBN lattice (according to the increase of lattice parameter, the degree of doping is about 0.025 at%; see below). Silicon substitutes boron, which leads to the formation of n-type semiconductor crystals [33] and, apparently, prevents the incorporation of magnesium atoms. It should be noted that cBN crystallization of is not observed at concentrations of silicon nitride higher than those necessary for the complete binding of magnesium nitride to siliconitride $MgSiN_2$.

During cBN crystallization from supercritical ammonia in the presence of silicon or silicon nitride, dark brown powder of doped cBN is formed, whose particles consist of aggregates of fine lamellar grains (Fig. 2c), and the maximum value of the lattice parameter $a$ is 3.6176 Å. The doping level was estimated in the framework of the ideal mixing model of isostructural cBN and hypothetical sphalerite-type SiN ($a$ = 4.299 Å when assuming that the length of Si–N bond is 1.862 Å, as in the case of α-$Si_3N_4$ [34]) and showed that the fraction of silicon in cBN structure can reach 0.19 at% (Fig. 3b). This value is comparable to the previously estimated solubility of silicon nitride in cBN at 7.7 GPa and temperatures above 1750°C [35].

### cBN doped with beryllium

As in the case of silicon, the incorporation of beryllium into cBN lattice should prevent the incorporation of magnesium. The degree of doping of cBN powders obtained from melts of the Mg–B–N system in the presence of beryllium compounds is low (about 0.06 at% according to estimates from an increase in the crystal lattice parameter; see below). However, the powders are light blue in color and electrically conductive, which clearly indicates that beryllium (or beryllium nitride) is soluble in cubic boron nitride. It should be noted that dark blue electrically conductive cBN crystals with beryllium content of about 0.1 at% are formed under similar conditions when BeO is used as additive.

During the synthesis from BN solutions in supercritical ammonia in the presence of beryllium nitride, the lattice parameter of the resulting cubic boron nitride (Fig. 2d) increases to 3.6166(2) Å. The amount of beryllium incorporated into cBN lattice was estimated using Vegard's law assuming the formation of cubic "solid solutions" $Be_xB_{1-x}N$ (Fig. 3c). For the hypothetical sphalerite-type BeN, the lattice parameter was taken equal to 4.005 Å, which corresponds to Be–N bond length of 1.734 Å in $Be_3N_2$ [36]. Thus estimated, the limiting solubility of beryllium in cBN upon crystallization from BN solutions in supercritical ammonia is about 0.21 at%. It should be noted that the formation of cBN was not observed in the studied pressure-temperature range when beryllium nitride concentration was more than 0.5 wt%.

Blue cBN crystals with a yellow-green tint having 1-2 orders of magnitude lower electrical conductivity compared to the crystals obtained in the presence of $Be_3N_2$ are formed when beryllium oxide is used as a dopant, which may indicate the incorporation of beryllium into cBN lattice in the form of BeO grouping (in the same way as in the absence of a nitrogen getter boron is incorporated into diamond in the form of BN grouping and forms boron-doped crystals with very low electrical conductivity [37]). In this case, the level of doping was estimated using Vegard's law for isostructural cBN and hypothetical sphalerite-type BeO (Be–O bond length 1.674 Å [38], and lattice parameter 3.866 Å). The maximum content of beryllium introduced in the form of BeO grouping is about 0.2 at%, which is the same as in the case of doping with beryllium nitride.

### *ESR spectra of doped cBN*

The ESR spectra of doped cBN samples (Fig. 4) show rather broad resonance lines due to averaging over all possible orientations of the particles of the studied micropowders. For the entire range of studied powders of doped cBN, four types of ESR-active impurities were registered, which correspond to line A ($g = 2.007$; $\Delta H \approx 10$ G), line B ($g = 2.002$; $\Delta H \approx 60$ G), line C ($g = 2.23$; $\Delta H \approx 330$ G), and line D ($g = 2.43$; $\Delta H \approx 1000$ G).

In the sample obtained by crystallization from a melt of the hBN–MgB$_2$ system (sample No. 26, see Table), the following three lines were recorded: narrow line A ($g = 2.007$; $\Delta H = 60$ G) characteristic for synthesis from melts of the Mg–B–N system, broad line D ($g = 2.43$; $\Delta H = 1000$ G), and also line C ($g = 2.24$; $\Delta H = 350$ G). The sample of cBN synthesized from a melt of the hBN–Mg$_3$N$_2$ system has a similar ESR spectrum. In the ESR spectra of samples 54 and 55, narrow line A is absent.

In sample No. 48 with sulfur addition, one broad line C ($g = 2.24$; $\Delta H = 360$ G) is present, and the signal is close to the noise level.

In the ESR spectra of samples 50 and 52 doped with silicon, only C line ($g = 2.23$; $\Delta H = 330$ G) is present.

All samples doped with beryllium give ESR line C ($g = 2.22$; $\Delta H = 300$ G), and the samples synthesized from a melt of the Mg–B–N system with addition of beryllium oxide additionally show narrow line A ($g = 2.007$ , $\Delta H = 10$ G).

The lack of resolution of broad lines in the ESR spectra does not allow an unambiguous interpretation of the results obtained. According to [39], line A is attributed to the association of defects based on nitrogen vacancies and interstitial atoms. Line B corresponds to the "charged nitrogen vacancy – unpaired electron" pair [40]. According to [41], line C—which is present in all samples—is apparently associated with carbon atoms that substitute nitrogen atoms, and line D may be explained by the fact that cBN structure contains magnesium atoms that substitute boron atoms.

*Semiconductor properties of doped cBN*

While magnesium doping is not accompanied by appearance of electrical conductivity in the formed crystals, the incorporation of silicon and beryllium atoms into cBN structure is accompanied by appearance of semiconducting properties. Doping with beryllium leads to p-type conductivity, while doping with silicon results in n-type conductivity [5]. In both cases, the electrical conductivity of single crystals (about 100 μm, samples 45 and 53) is around 0.001 $\Omega^{-1}$.

The dependences of electrical conductivity (σ) of silicon-doped cBN sinters on temperature at constant voltage and on applied voltage at constant temperature are shown in Figs. 5 and 6, respectively. The dependences of electrical conductivity of beryllium-doped cBN show similar behavior.

The observed lg σ(1/$T$) at U = const and lg σ($U^{½}$) at T = const dependences are close to linear, which indicates the Frenkel conduction mechanism, i.e. the lowering of the potential barrier of a Coulomb trap in a strong electric field. The temperature dependence of electrical resistivity is practically the same for the cBN(Si) and cBN(Be) sinters and is well reproducible upon repeated thermal cycling in the range of 150–650°C. Partial hydration of the samples occurs at temperatures below 150°C, especially in a humid atmosphere at room temperature. Electrical resistivity of a sinter heated to 600°C and quickly cooled down to room temperature decreased by three orders of magnitude within 24 hours at 20°C and relative air humidity 70%, and a hysteresis appears on the temperature dependence of electrical resistivity, which disappears upon subsequent heating to 300°C, after which the temperature dependence of electrical resistivity becomes reproducible upon repeated thermal cycling in the temperature range of 150–650°C at a water vapor partial pressure below 10 kPa. It should be noted that experiments performed in vacuum and in dry air (dried over $P_2O_5$) show complete reproducibility of results in the 20-650°C temperature range (dashed line in Fig. 5), which unambiguously indicates the sensitivity of the samples to moisture at temperatures below 150°C.

With a temperature increase in humid (70% at 20°C) air, conductivity hysteresis is observed i.e. a stable mode of high-temperature conductivity occurs only at 300°C, and the activation energy of conductivity changes from $E_{A1}$ = 0.7–0.8 eV at temperatures above 300°C to $E_{A2}$ = 0.35–0.40 eV below 300°C. Close values of activation energy were observed earlier in [42,43]. Apparently, such behavior is associated with the presence of the following two types of traps at the grain boundaries of the sinter: $OH^-$ and $O^{2-}$. Above 300°C, the first type of traps turns into the second type as a result of dehydration. Estimation of the grain boundary thickness by the method described earlier [44,45] gives a value of 5 Å, which corresponds to a double layer of oxygen ions. These grain boundaries increase the electrical resistivity of the sinters by six orders of magnitude.

Despite the high electrical resistivity in a dry atmosphere ($10^{10}$–$10^{11}$ Ω·m at 20°C and $10^6$ Ω·m at 600°C), sinters of semiconducting boron nitride can be used as thermosensitive elements of electronic device sensors in the temperature range of 150–650°C.

## Conclusions

There is negligible incorporation of magnesium into cubic boron nitride under the conditions of classical synthesis from a melt of the Mg–B–N system with quasi-equilibrium crystal growth, which is apparently due to the fact that the atomic radius of magnesium is much larger than the atomic radius of boron. However, the incorporation of magnesium into cBN lattice becomes possible by the synthesis from supercritical ammonia. Doping of cBN with silicon and beryllium is observed in both cases, and the degree of doping increases substantially in the case of crystallization of cubic boron nitride from BN solutions in supercritical ammonia at 3.9–4.4 GPa and 1100°C in comparison with the classical synthesis from melts of the Mg–B–N system at 4.2 GPa and 1400°C. Doping of cubic boron nitride with silicon and beryllium leads to the appearance of semiconducting properties, while doping with magnesium is not accompanied by the appearance of electrical conductivity of the formed cBN crystals.

## Acknowledgments

The authors are grateful to A.O. Semenkovich (VNIISIMS) for his assistance in conducting ESR measurements.

## References

1.  Wentorf H.R.  Cubic form of boron nitride.  *Chem. Phys.* 1957. Vol. 26, no. 4. P. 956-960.

2.  Rapoport E.  Cubic boron nitride: a review.  *Ann. Chim.* 1985. Vol. 10, no. 7. P. 607-638.

3.  Kurakevych, O.O., Solozhenko V.L.  High-pressure design of advanced BN-based materials. *Molecules* 2016. Vol. 21, no. 10. 1399.

4.  Chrenko R.M.  Ultraviolet and infrared spectra of cubic boron nitride. *Solid State Comm.* 1974. Vol. 14, no. 6.  P. 511-515.

5.  Wentorf R.H.  Preparation of semiconducting cubic boron nitride. *J. Chem. Phys.*  1962. Vol. 36, no. 8. P. 1990-1991.

6.  Shipilo V.B., Shishonok E.M., Akimov A.I. et al.  Studies of dielectric-properties of cubic boron-nitride.  *Dokl. Akad. Nauk Belarusi* 1985.  Vol. 29, no. 7. P. 604-606.

7.  Lukomskii A.I., Shipilo V.B., Shishonok E.M. et al.  Raman scattering of cubic boron nitride. *Phys. Status Solidi A* 1987. Vol. 102, no. 2. P. K137-K139.

8.  Taniguchi T., Teraji T., Koizumi S. et al.  Appearance of n-type semiconducting properties of cBN single crystals grown at high pressure. *Japan. J. Appl. Phys. Part 2 Letters* 2002. Vol. 41, no. 2A. P. L109-L111.

9.  Sachdev H.  Influence of impurities on the morphology and Raman spectra of cubic boron nitride. *Diamond Relat. Mater.* 2003. Vol. 12, no. 8. P. 1275-1286.

10. Shishonok E.M., Taniguchi T., Watanabe K. et al.  Low-frequency Raman scattering of Be-doped cubic boron nitride. *Diamond Relat. Mater.* 2003. Vol. 12, no. 3-7. P. 1133-1137.

11. Watanabe K., Taniguchi T., Kanda H. et al.  Polarized Raman scattering of impurity modes in beryllium-doped cubic boron nitride single crystals. *Appl. Phys. Lett.* 2003. Vol. 82, no. 18. P. 2972-2974.

12. Taniguchi T., Watanabe K., Koizumi S.  Defect characterization of cBN single crystals grown under HP/HT. *Phys. Status Solidi A* 2004. Vol. 201, no. 11. P. 2573-2577.

13. DeVries R.C., Fleischer J.F.  Phase equilibria pertinent to the growth of cubic boron nitride. *J. Crys. Growth* 1972. Vol. 13/14, P. 88-92.

14. Endo T., Fukunga O., Iwata M.  Precipitation mechanism of boron nitride in the ternary system of B-Mg-N.  *J. Mater. Sci.* 1979. Vol. 14, no. 7. P. 1676-1680.

15. Nakano S., Ikawa H., Fukunaga O.  High pressure reactions and formation mechanism of cubic boron nitride in the system boron nitride-magnesium nitride. *Diamond Relat. Mater.* 1993. Vol. 2, no. 8. P. 1168-1174.

16. H. Lorenz, I. Orgzall, E. Hinze  Rapid formation of cubic boron nitride in the system $Mg_3N_2$-hBN, *Diamond Relat. Mater.* 1995. Vol. 4, no. 8. P. 1050-1055.

17. Solozhenko V.L., Turkevich V.Z., Holzapfel W.B. On nucleation of cubic boron nitride in the BN-MgB$_2$ system. *J. Phys. Chem. B* 1999. Vol. 103, no. 38. P. 8137-8140.

18. Turkevich V.Z., Solozhenko V.L., Kulik O.G., Itsenko P.P., Sokolov A.N., Lutsenko A.N., Vashchenko A.N. Phase diagram of the Mg–B–N system at high pressures. *J. Superhard Mater.* 2003. Vol. 25, no. 6. P. 18-25.

19. Solozhenko V.L., Mukhanov V.A., Novikov N.V. On the p,T-region of cubic boron nitride formation. *Dokl. Akad. Nauk SSSR*, 1989. Vol. 308, no. 1. P. 131-133.

20. Solozhenko V.L. New concept of BN phase diagram: an applied aspect. *Diamond Relat. Mater.* 1994. Vol. 4, no. 1. P. 1-4.

21. Solozhenko V.L., Le Godec Y., Klotz S., Mezouar M., Turkevich V.Z., Besson J.-M. In-situ studies of boron nitride crystallization from BN solutions in supercritical N-H fluid at high pressures and temperatures. *Phys. Chem. Chem. Phys,* 2002. Vol. 4, no. 21. P. 5386-5393.

22. Solozhenko V.L., Solozhenko E.G., Petitet J.P. On cubic boron nitride crystals grown in the BN-N-H system. *J. Superhard Mater.* 2003. Vol. 25, no.1. P. 78-79.

23. Solozhenko V.L., Turkevich V.Z. Novikov N.V., Petitet J.P. On the cubic boron nitride crystallization in fluid systems. *Phys. Chem. Chem. Phys.* 2004. Vol. 6, no. 14. P. 3900-3902.

24. Solozhenko V.L. On the phase diagram of boron nitride. *Dokl. Akad. Nauk SSSR* 1988. Vol. 301, no.1. P. 147-149.

25. Solozhenko V.L. Boron nitride phase diagram. State of the art. *High Pressure Res.* 1995. Vol. 13, no. 4. P. 199-214.

26. Solozhenko V.L., Turkevich V.Z., Holzapfel W.B. Refined phase diagram of boron nitride. *J. Phys. Chem. B* 1999. Vol. 103, no. 15. P. 2903-2905.

27. Turkevich V.Z., Solozhenko V.L. Thermodynamic analysis of the nitrogen – hydrogen system at high pressures and temperatures. *J. Superhard Mater.* 2003. Vol. 25, no. 2. P. 11-13.

28. Khvostantsev L.G., Slesarev V.N., Brazhkin V.V. Toroid type high-pressure device: history and prospects. *High Pressure Res.* 2004. Vol. 24, no. 3. P. 371-383.

29. Mukhanov V.A., Sokolov P.S., Solozhenko V.L. On melting of B$_4$C boron carbide under pressure. *J. Superhard Mater.* 2012. Vol. 34, no. 3. P. 211–213.

30. Evain M. U-fit program. *Internal Report, I.M.N.* Nantes, France, 1992

31. Kurdyumov A.V., Solozhenko V.L., Zelyavski W.B. Lattice parameters of boron nitride polymorphous modifications as a function of their crystal-structure perfection. *J. Appl. Crystallogr.* 1995. Vol. 28. P. 540-545.

32. Reckeweg O., DiSalvo F.J. About binary and ternary alkaline earth metal nitrides. *Z. Anorg. Allgem. Chem.* 2001. Vol. 627, no. 3. P. 371-377.

33. Hirama, K. Taniyasu, Y., Yamamoto, H., Kumakura, K. Control of n-type electrical conductivity for cubic boron nitride (c-BN) epitaxial layers by Si doping. *Appl. Phys. Lett.* 2020. Vol. 116, no. 16. P. 162104.

34. Marchand R., Laurent, Y., Lang, J. Structure du nitrure de silicium alpha. *Acta Crystallogr. B* 1969. Vol. 25, no. 10. P. 2157-2160.

35. Gromyko S.N., Zalyavskii V.B., Kurdyumov A.V. et al. Solubility of silicon nitride in cubic boron nitride. *Dokl. Akad. Nauk SSSR*. 1989. Vol. 309, no. 5. P. 1115–1117.

36. Eckerlin P., Rabenau A. Die Struktur einer neuen Modifikation von $Be_3N_2$. *Z. Anorg. Allgem. Chem.* 1960. Vol. 304, no. 3-4. P. 218-229.

37. Laptev V.A., Pomchalov A.V. Growth and properties of dielectric and semiconductor single crystals of diamond. – *In: Synthesis of Minerals, VNIISIMS, Alexandrov*, 2000. Vol. 3. P. 75-143.

38. Sabine T.M., Hogg S. The wurtzite Z parameter for beryllium oxide and zinc oxide. *Acta Crystallogr. B* 1969. Vol. 25, no. 11. P. 2254-2256.

39. Shipilo V.B., Rud' A.E., Dutov A.G. et al. Radiation defects of sphalerite boron nitride. *Izv. Akad. Nauk SSSR, Neorg. Mater*. 1991. Vol. 27, no. 8. P. 1637-1640.

40. Khusidman M.B., Neshpor V.S. On F-centers in hexagonal boron nitride enriched with [10]B isotope. *Fiz. Tverd. Tela*, 1968. Vol. 10, no. 4. P. 1229-1231.

41. Mukhanov V.A., Semenkovich A.O. ESR analysis of powders of doped cubic boron nitride. *In: Trudy VNIISIMS, Alexandrov* 1997. Vol. 14. P. 278-280.

42. Yin, H., Pongrac I., Ziemann P. Electronic transport in heavily Si doped cubic boron nitride films epitaxially grown on diamond (001). *J. Appl. Phys.* 2008. Vol. 104, no. 2. 023703.

43. Li X., Feng S., Liu X et al. Investigation on cubic boron nitride crystals doped with Si by high temperature thermal diffusion. *Appl. Surf. Sci.* 2014. Vol. 308. P. 31-37.

44. Valeev K.S., Kvaskov V.B. *Nonmetallic Metal-Oxide Semiconductors*. Moscow: Energoizdat, 1983. 160 p.

45. Kvaskov V.B. *Semiconductor Devices with Bipolar Conductivity*. Moscow: Energoatomizdat, 1988. 128 p.

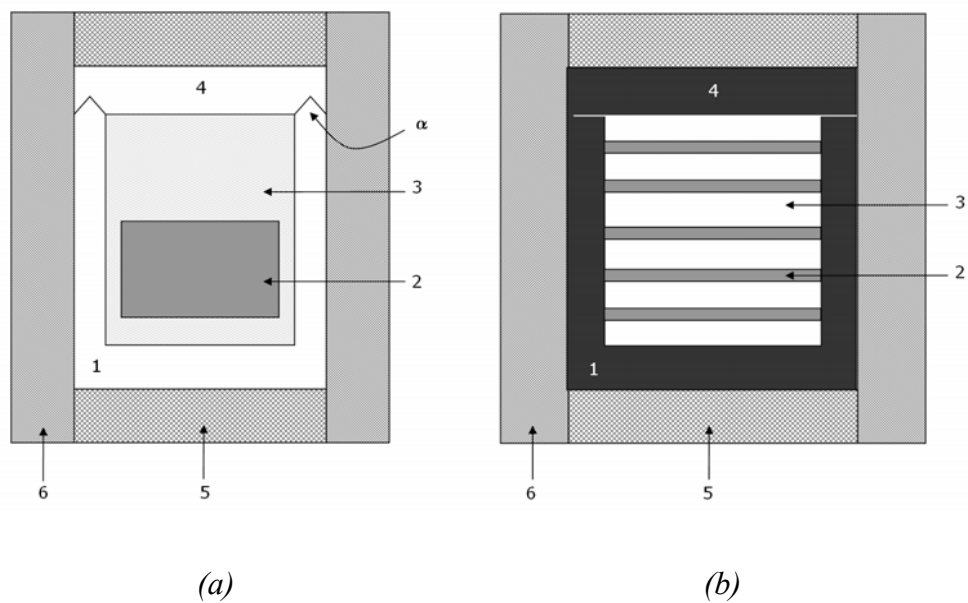

*(a)*                    *(b)*

Fig. 1   (*a*) Reaction cell for cBN synthesis by crystallization from solution of boron nitride in supercritical ammonia: (*1*) copper capsule (angle α = 90°); (*2*) a mixture of hBN, magnesium boride (or magnesium boron nitride) and a dopant; (*3*) condensed ammonia; (*4*) copper capsule cap; (*5*) composite electrical leads; and (*6*) gasket made of lithographic stone.  (*b*) Reaction cell for cBN synthesis by crystallization from melts of the Mg–B–N system: (*1*) graphite heater; (*2*) layer of a mixture of magnesium boride (or magnesium boron nitride) and a dopant; (*3*) layer of hBN; (*4*) graphite cap; (*5*) composite electrical leads; and (6) gasket made of lithographic stone.

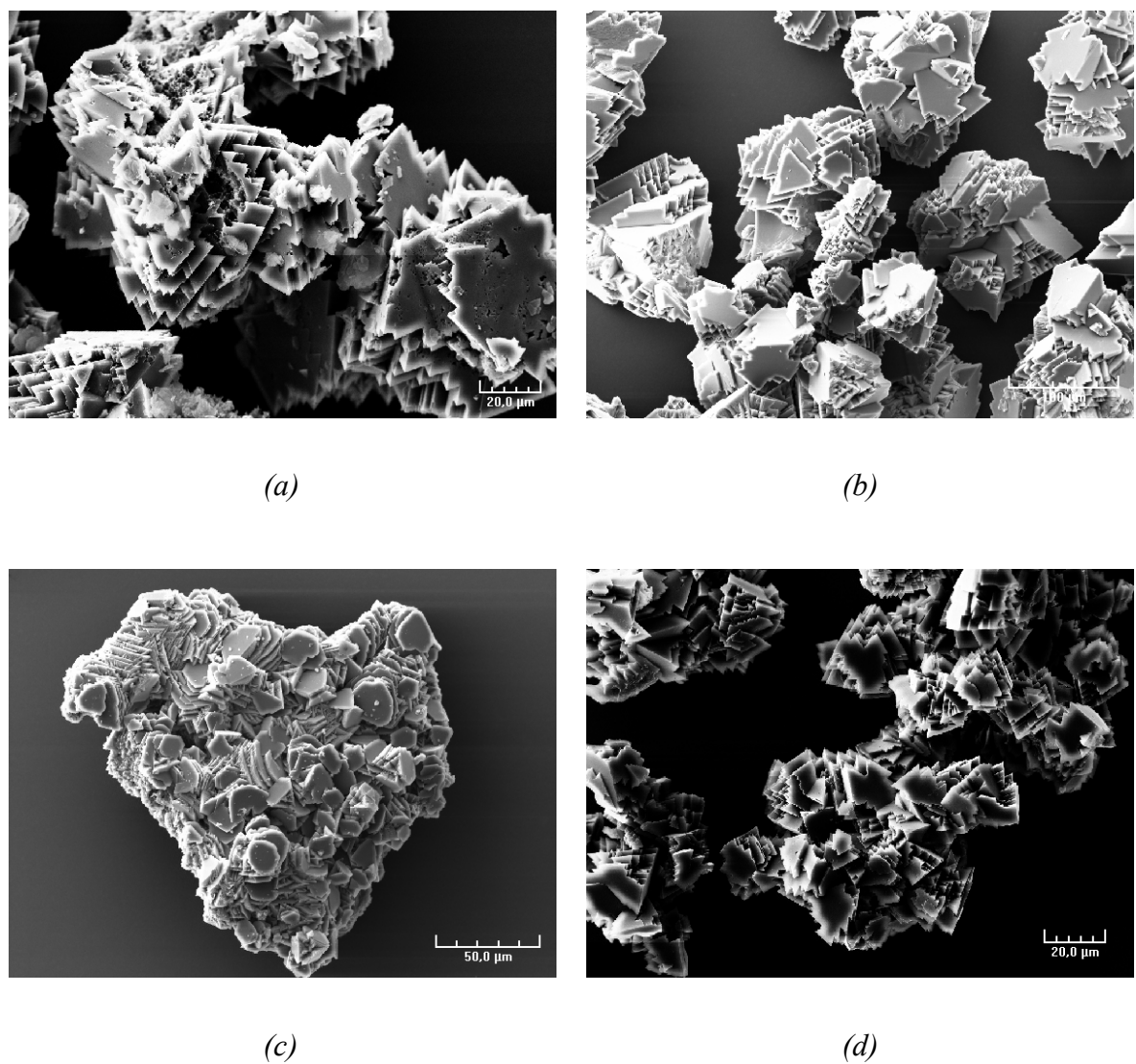

*(a)*  *(b)*

*(c)*  *(d)*

Fig. 2  Electron microscopic images of cubic boron nitride micropowders synthesized by crystallization from BN solutions in supercritical ammonia: (*a*) undoped, (*b*) "doped" with sulfur, (*c*) doped with silicon, (*d*) doped with beryllium.

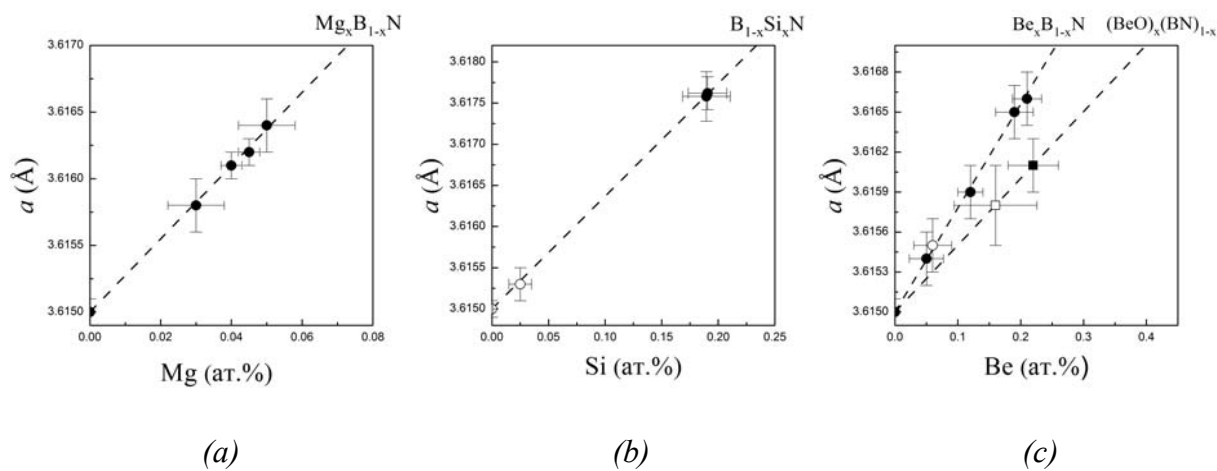

*(a)*            *(b)*            *(c)*

Fig. 3    Evaluation of the composition of cBN doped with (*a*) magnesium, (*b*) silicon and (*c*) beryllium using Vegard's law. Linear dependences of *a*-parameter of the cubic lattice on the dopant content correspond to ideal solid solutions of hypothetical sphalerite-type (*a*) MgN, (*b*) SiN, and (*c*) BeN and BeO in cubic boron nitride. The lattice parameters of cBN synthesized by crystallization from BN solutions in supercritical ammonia are represented by solid symbols, and obtained by traditional synthesis from melts of the Mg–B–N system are shown by open symbols.

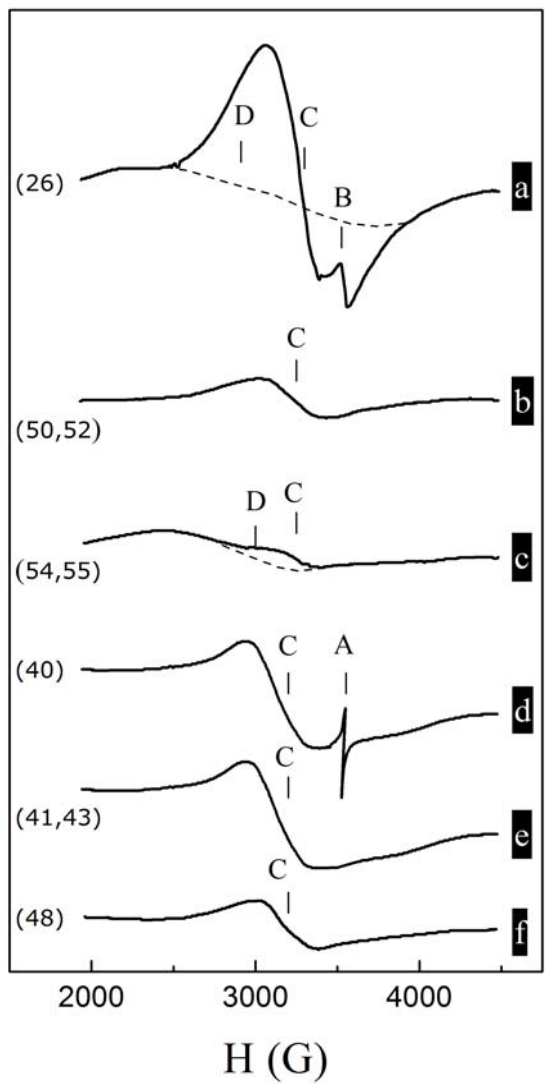

Fig. 4    ESR spectra of cBN powders: (a) undoped; (b) doped with silicon nitride, (c) doped with magnesium nitride, (d) doped with beryllium oxide, (e) doped with beryllium nitride, and (f) "doped" with sulfur. The undoped sample and the sample doped with beryllium oxide were obtained by traditional synthesis, and the rest of the samples were synthesized by crystallization from BN solutions in supercritical ammonia. Sample numbers are given in brackets (see Table).

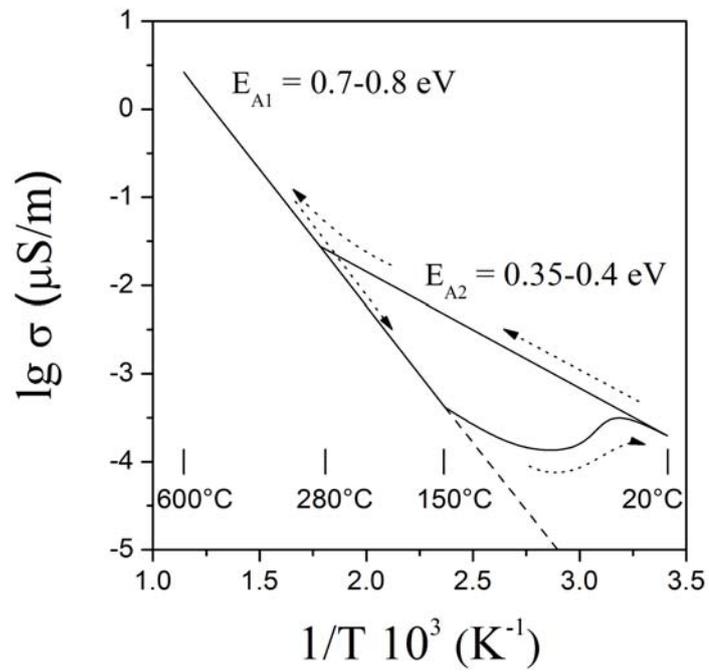

Fig. 5    Temperature dependences of the electrical conductivity of cBN sinters doped with silicon. The arrows indicate the direction of changing the parameters. The hysteresis is observed only in humid air, while it is absent when the measurements are conducted in dry air and in vacuum.

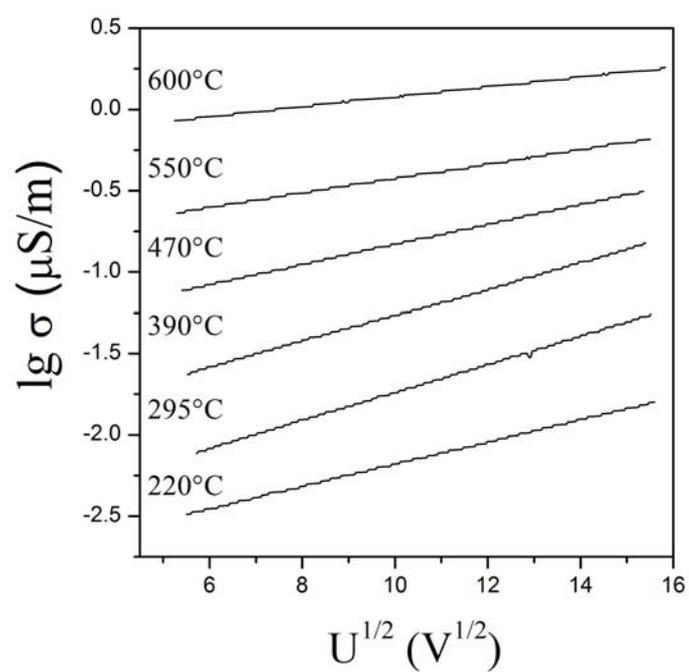

Fig. 6    Dependence of electrical conductivity on voltage for cBN sinters doped
with silicon.

Table      Synthesis conditions, lattice parameters, and degrees of doping of cubic boron nitride obtained by various methods

| Sample number | Reaction mixture composition, wt.% | Pressure and temperature | $a$, Å | Doping element, ат.% |
|---|---|---|---|---|
| Classical synthesis from melts of the Mg–B–N system | | | | |
| 26 | 15% MgB$_2$; 85% hBN | 4.2 GPa; 1400°C | 3.6150(1) | Mg, 0% |
| 40 | 20% (17% BeO, 83% Mg$_3$N$_2$); 80% hBN | 4.2 GPa; 1400°C | 3.6158(3) | Be (+O), 0.016% Be (+N), 0.010% |
| 45 | 20% (15% Be$_3$N$_2$, 85% Mg$_3$N$_2$); 80% hBN | 4.2 GPa; 1400°C | 3.6155(2) | Be (+N), 0.06% |
| 53 | 20% (10% Si$_3$N$_4$, 90% Mg$_3$N$_2$); 80% BN | 4.2 GPa; 1400°C | 3.6153(2) | Si, 0.025% |
| Synthesis from solutions of boron nitride and magnesium compounds in supercritical ammonia | | | | |
| 54 | 35% NH$_3$; 5.2% MgB$_2$; 59.8% hBN | 4.4 GPa; 1100°C | 3.6161 | Mg, 0.04% |
| 46 | 50% NH$_3$; 5% Mg$_3$N$_2$; 45% hBN | 4.0 GPa; 1100°C | 3.6162 | Mg, 0.045% |
| 55 | 50% NH$_3$; 10% Mg$_3$N$_2$; 40% hBN | 4.0 GPa; 1100°C | 3.6164(2) | Mg, 0.05% |
| 48 | 0.0795% S; 49.07% NH$_3$; 7.63% MgB$_2$; 43.22% hBN | 3.9 GPa; 1000°C | 3.6158(2) | Mg, 0.03% |
| 50 | 3.12% Si; 37.51% NH$_3$; 9.37% MgB$_2$; 50% hBN | 4,2 GPa; 1100°C | 3,6176(2) | Si, 0.19% |
| 52 | 7.12% Si$_3$N$_4$; 48.4% NH$_3$; 10.32% MgB$_2$; 34,17% hBN | 4,2 GPa; 1100°C | 3,6176(3) | Si, 0.19% |
| 41 | 0.11% Be$_3$N$_2$; 40.88% NH$_3$; 5.39% MgB$_2$; 0.39% Mg$_3$(BN$_2$)$_2$; 53.05% hBN | 4.2 GPa; 1100°C | 3.6154(2) | Be, 0,05% |

| 43 | 0.23% $Be_3N_2$; 35.87% $NH_3$; 5.3% $MgB_2$; 1.25% $Mg_3(BN_2)_2$; 57.35% hBN | 4.2 GPa; 1100°C | 3.6159(2) | Be (+N), 0.12% |
|---|---|---|---|---|
| 47 | 0.369% $Be_3N_2$; 36.36% $NH_3$; 5.68% $MgB_2$; 1.6% $Mg_3(BN_2)_2$; 56.99% hBN | 4.2 GPa; 1100°C | 3.6165(2) | Be (+N), 0.19% |
| 42 | 0.42% $Be_3N_2$; 34.24% $NH_3$; 5.28% $MgB_2$; 60.06% hBN | 4.0 GPa; 1100°C | 3.6166(2) | Be (+N), 0.21% |
| 36 | 1.33% BeO; 35.10% $NH_3$; 6.52% $MgB_2$; 57.03% hBN | 4.2 GPa; 1100°C | 3.6161(2) | Be (+O), 0.022% Be (+N), 0.014% |
| 39 | 1.31% BeO; 36.92% $NH_3$; 6.44% $MgB_2$; 55.33% hBN | 4.2 GPa; 1100°C | 3,6161(2) | Be (+O), 0.022% Be (+N), 0.014% |